\documentclass[12pt]{book}
\usepackage{plenum,psfig}
\begin{document}
\chapter{Multifragmentation at Intermediate Energy: 
Dynamics or Statistics?}

\author{Luc Beaulieu, Larry Phair, Luciano G. Moretto, 
and Gordon J. Wozniak}

\affiliation{Nuclear Science Division\\
Lawrence Berkeley National Laboratory\\
Berkeley, California 94720\\
\\
Presented at the 14$^{th}$ Winter Workshop\\ 
on Nuclear Dynamics, Snowbird, Utah,\\ 
January 31- February 7, 1998}

\section{Introduction}

Since the observation of a power-law behaviour in the charge 
distributions, characteristic of critical phenomena~\refnote{\cite{Fis67,Stu92}}, 
in proton induced reactions at relativistic energies, the production of 
multiple intermediate mass fragments (IMF)~\refnote{\cite{Bor92,Mor93}}, 
typically $3\le Z \le 20$, has been touted as a signature of the nuclear 
liquid-gas phase transition~\refnote{\cite{Fin82,Sie83,Pan84}}. While this may
be the case in peripheral reactions e.g. projectile or spectator 
breakup~\refnote{\cite{Des93,Kre93,Gil94,Poc95,Ben95,Bea96c,Mas96,Bea96b,Sch96,
Poc97,Hau98}}, the situation becomes less clear when one looks at more 
central reactions. In particular, it has been shown that the dissipative 
binary mechanism~\refnote{\cite{Lot92,Que93,Lec94,Pet95,Bea96}} contributes 
95\% or more of the reaction cross section~\refnote{\cite{Pet95,Bea96}}. Yet, 
as long as the sources are thermalized, it has been shown that a 
characteristic signature for phase coexistence can be extracted from the charge 
distributions~\refnote{\cite{Pha95,Mor95b}}. The situation 
is further complicated by the experimental observation of a significant 
contribution to the fragment yields from a third source formed between 
the projectile and target~\refnote{\cite{Monto94,Luka97,Laro97,Tok95,Leco95,Demp96}}. 
Most of these observations were made using velocity plots (see for example 
ref.~\cite{Luka97}) which are useful in assigning a given particle to its 
primary source. This evidence points out the importance of dynamics in the 
entrance channel. Unfortunately, it tells very little about the intrinsic 
properties of the sources themselves. In particular, it does not disclosed the nature 
of the fragmentation process producing the detected ``cold'' IMF, i.e. at 
$t \rightarrow \infty$.

In the following, we will consider two contradictory claims that have been
advanced recently: 1) the claim for a predominantly dynamical fragment
production mechanism~\refnote{\cite{Tok96,Tok97}}; and 2) the claim for a dominant 
statistical and thermal process~\refnote{\cite{Mor95,Tso95,Mor97,Pha96,Bea98,Mor93b}}. 
We will present a new analysis in terms of Poissonian reducibility and thermal 
scaling, which addresses some of the criticisms of the binomial 
analysis~\refnote{\cite{Tok97,Tsa97,Sku97}}.

\section{Dynamical fragment production}

To make a statement about the nature and mechanism of fragmentation, it is 
necessary to probe directly any competition, or lack thereof, between the 
emission of various particle species as a function of excitation energy. The 
task is then to find a global observable that best follows the increase in 
excitation energy or dissipated energy. IMF multiplicity, $N_{IMF}$, 
and total transverse energy, $E_{t}$ have both been used to infer a decoupling 
between light charged particles (LCP) and IMF production~\refnote{\cite{Tok96,Tok97}}.

\subsection{$N_{IMF}$ as global observable}

Recently, it was claimed by Toke et al.~\refnote{\cite{Tok96}} that IMF 
production is predominantly a dynamical process. The evidence came by 
looking at different particle multiplicities, and their corresponding 
transverse energies, as a function of IMF multiplicity.

\begin{figure}[t]
\centerline{\psfig{file=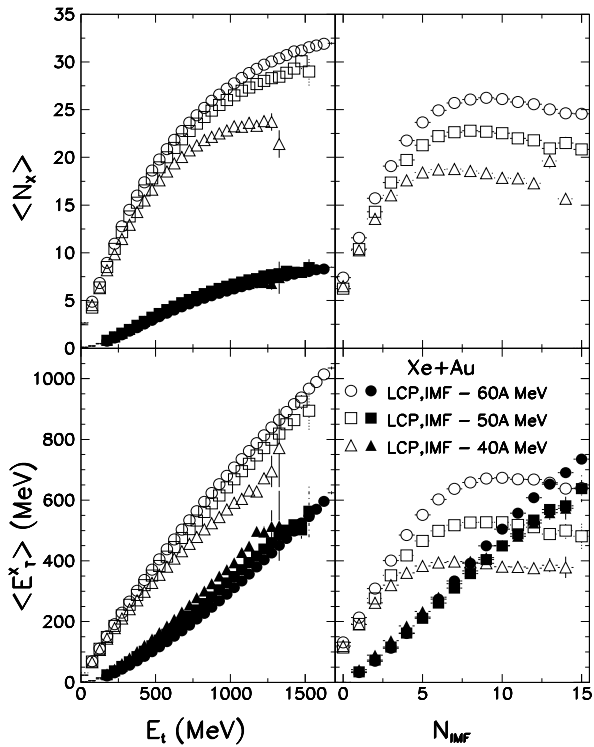,height=10.0cm,width=8.0cm}}
\caption{Average LCP multiplicity (upper panels), IMF multiplicity (upper
left) and average transverse energy associated to LCP and IMF (bottom 
panels) as a function of the total transverse energy (left panels) and
IMF multiplicity (right panels) for the Xe + Au reaction 
between 40 and 60 MeV/nucleon.}
\end{figure}

The argument was made as follows. The multiplicities of neutrons ($N_{n}$) 
and light charged particles ($N_{LCP}$), represent a good measurement of the 
thermal excitation energy of the system, $E^{*}$, as do the transverse energies
of the LCPs, $E_{t}^{LCP}$. Using $N_{IMF}$ as a global variable, a 
fast and simultaneous saturation of $N_{n}$, $N_{LCP}$ and $E_{t}^{LCP}$
was observed in the reaction Xe+Bi at 28A MeV~\refnote{\cite{Tok96}}. 
This saturation occurs around $N_{IMF}$ = 2-3. The authors conclude that, 
since most of the IMFs (up to 12) are produced after the saturation, 
there is a ``critical'' excitation energy above which the IMFs are produced
without competing with the LCPs. This apparent decoupling of IMF production 
from that of the LCPs is interpreted as due to the onset of a dynamical process.

We have explored the above behaviour in a systematic study of 
Xe+Au reactions (similar to Xe+Bi) at 40A, 50A and 60A MeV. The data 
were taken in two different experiments at the NSCL using the MSU Miniball 
4$\pi$ array and the LBL forward array~\refnote{\cite{Sou92,Keh92}}. The right 
panels of Fig. 1 show the evolution of $N_{LCP}$, $E_{t}^{LCP}$ 
and $E_{t}^{IMF}$ as a function of $N_{IMF}$. The saturation 
is clearly present in both observables related to LCP. However, as the beam 
energy increases, the saturation point moves toward higher IMF mutiplicities. 
At 60A MeV, the saturation occurs at $N_{IMF} \sim$ 8. If one follows the 
interpretation mentioned above, one might be led to conclude 
that {\it most} of the IMF are produced {\it before} the saturation 
``critical energy'' and that therefore the IMF production might be 
statistical and thermal in nature. In any case, the features shown in 
Fig. 1 are sufficiently intriguing to warrant further study.

To do so, we have performed a simulation using the SMM
model~\refnote{\cite{Bon95}}. 
We have considered the breakup of Au nuclei with a triangular excitation 
energy ($E^*$) distribution ranging from 0.5A to 6.0A MeV (note that a flat 
distribution does not change the conclusion but that a triangular one 
is closer to the impact parameter weighted behaviour of the cross section). 
The maximum average number of IMFs for this simulation is about 4, similar 
to the Xe+Bi case. Cuts on $N_{IMF}$ were done and are shown in 
the right panels of Fig. 2. Here, as in the experiment, we notice a fast and 
simultaneous saturation of $N_{n}$, $N_{LCP}$ and $E_{t}^{LCP}$. However, the 
fragmentation process is, by the nature of the model, of statistical 
origin. Inspection of the figure reveals that saturation occurs around 
$N_{IMF}$=4, which corresponds to the maximum average number, 
$\left< N_{IMF} \right>_{max}$. In the model, the average value of 
$N_{IMF}$ increases with $E^*$ until $\left< N_{IMF} \right>_{max}$ is reached 
at $E^*_{max}$. Therefore, $N_{IMF}$ is, on average, a rough measure of 
excitation energy for $N_{IMF} < \left< N_{IMF} \right>_{max}$. For values of 
$N_{IMF} > \left< N_{IMF} \right>_{max}$, there is no increase of $E^*$. 

For a given $E^*$, the IMF distribution is characterized not only by its mean 
but also by its variance. Although $\left< N_{IMF} \right>_{max}$=4, Fig. 2 
(right panels) shows that events with up to 12 IMF are present. Cutting on 
$N_{IMF}$ past its average maximum value probes a {\it nearly constant} 
excitation energy. This is nicely illustrated by the saturation of neutron 
and LCP multiplicities, which are also sensitive to $E^*$. This is a general 
feature of any statistical model as pointed out by Phair et 
al.~\refnote{\cite{Pha98a}}. Note that the increase of $E_{t}^{IMF}$ with 
$N_{IMF}$ is due to the trivial autocorrelation between the two quantities. 

Returning to the data (Fig. 1, right panels), as the beam energy increases, 
the excitation energy and IMF production increase. Therefore, in a statistical 
picture, the change in the ``critical saturation energy'' is due to the 
increase of excitation energy (dissipated energy) with beam energy, and 
correspondingly, to an increase of $\left< N_{IMF} \right>_{max}$ with 
excitation energy. If, for a given reaction, IMF were produced dynamically,
why should the ``critical saturation energy'' change with beam energy?

\begin{figure}[t]
\centerline{\psfig{file=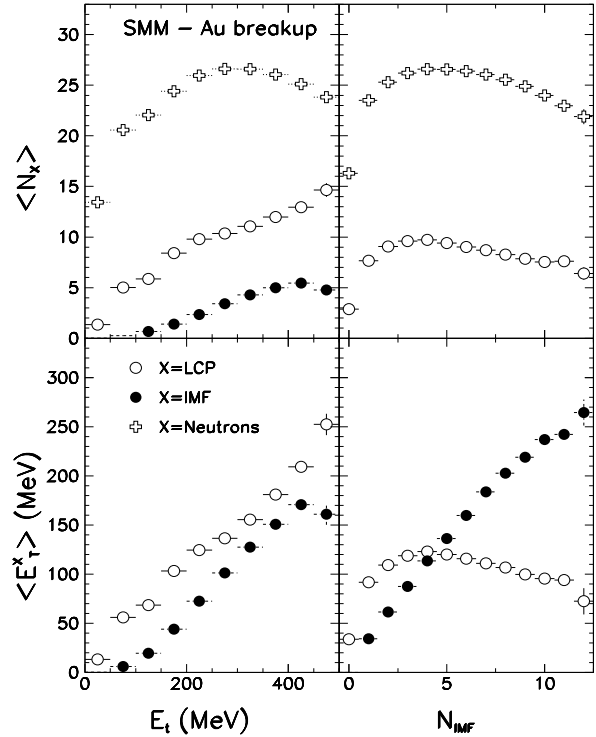,height=10.0cm,width=8.0cm}}
\caption{Average LCP and neutron multiplicities (upper panels), 
IMF multiplicity (upper left) and average transverse energy associated 
to LCP and IMF (bottom panels) as a function of the total transverse 
energy (left panels) and IMF multiplicity (right panels) for the breakup
of Au nuclei in the SMM model. For details see text.}
\end{figure}

\subsection{$E_t$ as global observable}

The same authors have suggested~\refnote{\cite{Tok97}} that the ``evidence'' 
for dynamical IMF production shown in the previous section might already be 
contained in the evolution of the same quantities ($N_{n}$, $N_{LCP}$, $E_{t}^{LCP}$, $N_{IMF}$ and $E_{t}^{IMF}$) as a function of the total 
transverse energy, $E_{t}$. Again, the authors have observed a fast and 
simultaneous saturation of $N_{n}$, $N_{lcp}$ and $E_{t}^{LCP}$ as a function 
of $E_t$, but a continuous increase of $E_{t}^{IMF}$ and $N_{IMF}$. In their 
work (ref.~\cite{Tok97}, Fig. 2), they state, correctly, that if $E_{t}$ were 
a good measure of excitation 
energy, and the IMF were produced statistically, such saturations should not 
occur. Indeed, our statistical simulation (left panels of Fig. 2) shows that 
$N_{LCP}$ and $E_{t}^{LCP}$ increase monotonically with $E_{t}$ and, at no 
point, is $E_{t}^{IMF}$ greater than $E_{t}^{LCP}$. Thus, the behaviour of the 
Xe+Bi experimental results, if correct, cannot be explained by statistical models.

In fact, saturations in $N_{LCP}$ and $E_{t}^{LCP}$ are not observed in 
comparable data for the Xe+Au reactions as shown in Fig. 1 (left panels). 
$N_{LCP}$ increases smoothly with $E_t$, as does $E_{t}^{LCP}$. Notice that 
$E_{t}^{LCP}$ is always larger than $E_{t}^{IMF}$. For the Xe+Au at 50A MeV, 
the ratio $E_{t}^{IMF}/E_{t}^{LCP}$ is always smaller than 0.3. This result
is strongly at variance with the Xe+Bi data where the IMF contribute
up to 80\% to the total $E_t$. The Xe+Bi data are also very different from 
preliminary results of the Xe+Au reaction at 30A MeV (sister reaction of the 
Xe+Bi at 28A MeV)~\refnote{\cite{Col98,Pha98b}}, whose behaviour is similar 
to the data at higher energies in Fig. 1. 

The dramatic difference between the Xe+Bi and the Xe+Au data may be due to 
the experimental set-up used for the former experiment, the Dwarf 
Ball~\refnote{\cite{Str90}}, whose detectors are made of thin, 4mm, 
CsI(Tl). Such thin detectors have a punch through energy of 30A MeV
for proton and alpha particles. While the thickness of these detectors
is suitable for fragments, they are too small to stop LCPs in this beam 
energy range. If the punch through effect is not properly corrected, the 
total kinetic energy associated to LCPs will be severely underestimated. This, 
by construction, leads to a much larger percentage of the transverse energy 
carried by the IMFs at a given total $E_t$. A detailed analysis of the Xe+Au 
systematic, and its comparison to Xe+Bi data, including a software replica of 
the Dwarf Ball, in under way~\refnote{\cite{Pha98b}}. However, it is already 
clear that the features presented in Fig. 2 of ref.~\cite{Tok97} are due to 
an experimental artifact, rather than to dynamical decay.

\section{Statistical fragment production}

Another way of approaching the fragmentation process is to rely on methods that
worked well at lower energies, and permitted the understanding of low energy 
particle evaporation and fission of a compound nucleus. At low energies, emission 
probabilities and excitation functions have been far more successful then kinematical 
variables at suggesting whether the process is statistical (compound
nucleus decay) or dynamical (direct reactions)~\refnote{\cite{Mor69}}.

The increase of fission probability as a function of excitation energy (directly
related to the temperature at low energies) can be cast in terms of a 
Boltzmann factor depending on the temperature and the fission barrier. 
The corresponding Arrhenius plots obtained from fission data are linear 
and cover a range from 2 to 6 order of magnitudes~\refnote{\cite{Mor69}}!

Recently, similar behaviour has been found in multifragmentation 
data~\refnote{\cite{Pha95,Mor95b,Mor95,Tso95,Mor97,Pha96,Bea98,Mor93b}}.
It has been shown that the probability $P_n$ of emitting $n$ intermediate 
mass fragments (IMFs) can be reduced to the probability of emitting a 
single fragment through the binomial equation~\refnote{\cite{Mor95,Tso95,Mor97}}. 
The extracted elementary emission probabilities $p$ were also shown to give 
linear Arrhenius plots when $\log 1/p$ is plotted vs $1/\sqrt{E_t}$.  In 
going from reducibility to thermal scaling, the only assumption needed 
is that $E_t$ is proportional to excitation energy (or temperature). 
We should therefore include a few words on $E_t$. From an experimental 
point of view, $E_t$ represents a measure of the total energy dissipated 
in the reaction. It can be written as follows
\begin{equation}
E_t = E_t^{pre-equilibrium}+E_t^{rotation}+E_t^{flow}+
E_t^{Coulomb}+E_t^{thermal}
\end{equation}

In other words, the thermal portion of $E_t$ is drowned in an ocean
of other contributions, as is the thermal excitation energy itself! 
For example, if we take the SMM model, and try to reproduce the 
$\left<N_{IMF}\right>_{max}$ of a given reaction, usually the $E_t$ (thermal 
$E_t$) range is too small by a factor of at least 2. However, the important 
unanswered question is, is $E_t$ tracking the increase of thermal excitation 
energy? We believed that it does but this remains to be proven.

In the hypothesis that the temperature $T$ is proportional to $\sqrt{E_t}$, 
these linear Arrhenius plots suggest that $p$ has the Boltzmann form 
$p\propto\exp(-B/T)$. This form holds 
for many different reactions from reverse to normal kinematics and almost 
over the complete intermediate energy range. Similarly, the charge 
distributions for each fragment multiplicity $n$ and the experimental 
particle-particle angular correlation are also both reducible to the 
distribution of individual fragments and 
thermally scalable~\refnote{\cite{Pha95,Mor95b,Pha96}}.

However, this approach has been meet with several criticisms. 
First, the binomial decomposition has been performed 
on the $Z$-integrated fragment multiplicities (IMF), typically associated 
with $3 \le Z \le 20$. Thus, the Arrhenius plot generated with the 
resulting one fragment probability $p$ is an average over a range of $Z$ 
values. A second ``problem'' lies in the transformation from the 
excitation $E^*$ to the transverse energy $E_t$. It was shown that if the 
width associated with this transformation is too large, than the linearity 
of the Arrhenius plots constructed with the elementary probability $p$ would 
be lost in the averaging process~\refnote{\cite{Tok97}}. While both binomial 
parameters $p$ and $m$ are individually susceptible to this problem, the 
product of the two, $\left< n \right>$ =  $\left< mp \right>$ has been 
shown to be very resilient to the averaging process~\refnote{\cite{Tok97}}. 
Finally, the fact that IMFs as a category can contribute a fair amount to 
$E_t$, about 30\% maximum for the Xe+Au reaction at 50A MeV, has been pointed 
to as a possible source of autocorrelation between $p$ and $\sqrt{E_t}$ 
leaving its interpretation questionable~\refnote{\cite{Tsa97,Sku97}}. 

In the following, we will present results from a new analysis~\refnote{\cite{Bea98}} 
in which we look for reducibility and thermal scaling at the level 
of individual fragments of charge $Z$, and, at the same time,
answer in a rather elegant way the above mentioned criticisms.

\begin{figure}[tp]
\centerline{\psfig{file=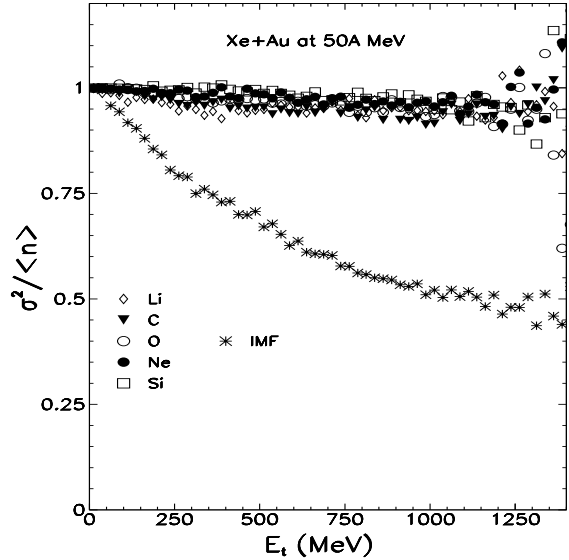,height=7.5cm,width=7.5cm}}
\caption{The ratio of the variance to the mean number of Li, C, O, 
Ne and Si fragments (open and solid symbols) emitted from the reaction 
$^{129}$Xe+$^{197}$Au at 50A MeV. The star symbols show the same 
ratio for all IMFs ($3\le Z\le 20$).}
\end{figure}

\subsection{Poissonian reducibility}

We analyze the fragment multiplicity distributions 
for each individual fragment $Z$ value. This restriction has the rather 
dramatic effect of decreasing the elementary probability $p$, compared to 
that associated with the total IMF value, to the point where the variance 
over the mean for any $Z$ is very close to one for all values of 
$E_t$~\refnote{\cite{Mor97,Bea98}}(Fig. 3). This means that the binomial 
distribution tends to its Poissonian limit. In this limit, the quantities 
$m$ and $p$ are not individually extracted, but it is rather the quantity 
$\left< n \right>$ =  $\left< mp \right>$ that is obtained.  The Poisson 
distribution is expressed as
\begin{equation}
P_n(Z)=\frac{\left< n_{Z}\right>^ne^{-\left< n_{Z}\right>}}{n!}
\end{equation}
where $n$ is the number of fragments of {\it a given $Z$} and the average value
$\left< n_{Z} \right>$ is a function of $E_t$. We can verify the ability of 
Eq. 1 to reproduce the n-fold probability distribution, $P_n$, for Li fragments
in Fig. 4 (left panel). The symbols are experimental n-fold probabilities, 
while the lines are the probabilities obtained by introducing the experimental 
average values in Eq. 1. For all the reactions studied, Poissonian 
fits (Eq. 1) were excellent for all $Z$ values starting from $Z$=3 up to 
$Z$=14 over the entire range of $E_t$~\refnote{\cite{Bea98}}. $\left< n_Z \right>$ is 
now the only quantity needed to describe the emission probabilities $P_n$ of 
charge $Z$. Thus we conclude that reducibility (now Poissonian reducibility) 
is verified at the level of individual $Z$ values for many different 
systems. Moreover, reducibility is tested for each ($Z$,$E_t$) 
combination. For example, in Fig. 4, the reducibility is tested 40 times just 
for $Z$=3. Reducibility, binomial or Poissonian, is an {\it experimental} 
observation, demonstrating that fragment emission is a stochastic process.

\begin{figure}[tp]
\centerline{\psfig{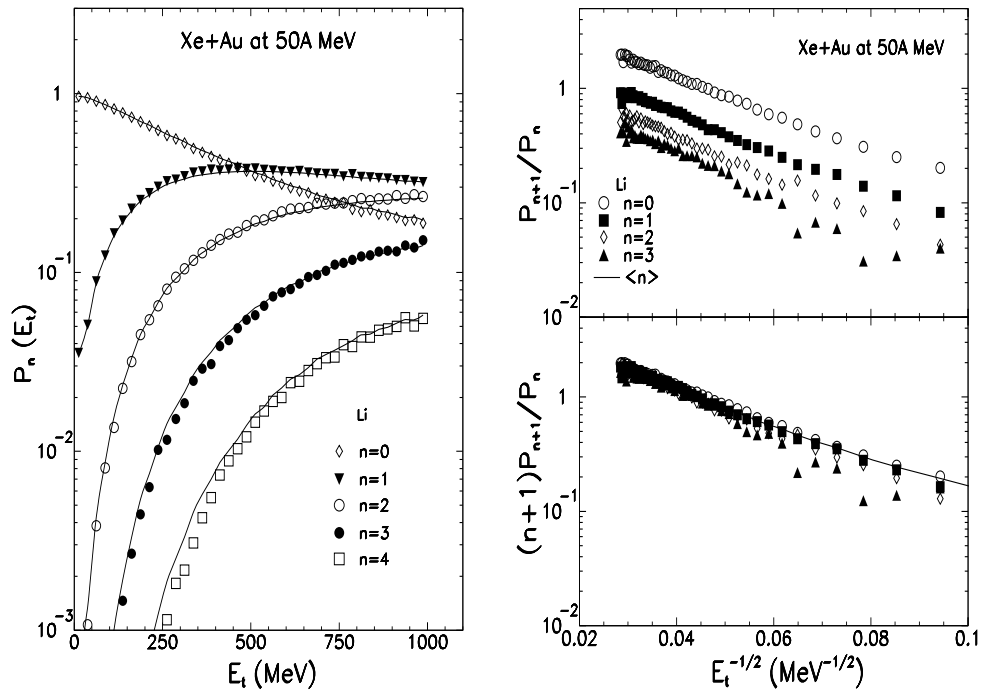}}
\caption{Left panel: The excitation functions $P_n$ for lithium 
emission from the reaction $^{129}$Xe+$^{197}$Au at 50 A MeV.
The lines are Poisson fits to data. The ratio of one n-fold to the next
is shown in the upper right panel and the appropriate scaling in
the lower right panel. The line in the lower right panel is the
experimental average number, $\left< n_Z \right>$.}
\end{figure}

\subsection{Thermal scaling}

In order to verify thermal scaling, we can first look at the ratio
of one fold to the next, $P_{n+1}/P_n$ as in ref.~\cite{Mor93b}.
The results yield linear plots versus $1/\sqrt{E_t}$ as shown in
Fig. 4 (right panel). However, these plots are not all independent;
in fact, from Eq. 1, one find that $P_{n+1}/P_n = \left<n\right>/n+1$.
Correcting the ratio by the trivial $n+1$ factor collapses all the curves
into a single one, which follows nicely the line of the experimental 
average values. Consequently, we generate Arrhenius plots by plotting 
directly $\log\left< n\right>$ vs $1/\sqrt{E_t}$. The left panel of Fig. 5 
gives a family of these plots for the Xe+Au reaction at 50A MeV, and 
$Z$ values extending from $Z$=3 to $Z$=14. These Arrhenius plots 
are strikingly linear over factors of 10 to 60, and their slopes increase 
smoothly with increasing $Z$ value. The overall linear trend demonstrates 
that thermal scaling is also present when individual fragments of a specific 
$Z$ are considered. 

The advantage of this procedure is readily apparent. 
For any given reaction, thermal scaling is verifiable for as many atomic
numbers as are experimentally accessible (12 in this case). Futhermore, to 
generate this figure, Poissonian reducibility has been tested 480 times. 
This is an extraordinary level of verification of 
the empirical reducibility and thermal scaling with the variable $E_t$. 

Additionally, as discussed above, $\left< n_Z \right>$ is free of any 
distortion due to averaging when going from $E^*$ to $E_t$~\refnote{\cite{Tok97}}. 
Also, because of the dominance of the zero fold probability, the 
average contribution of a particular $Z$ to $E_t$ is very small, $\le$ 5\%, 
thus minimising the risk of autocorrelation. Still, to be sure that 
there is no autocorrelation, we have repeated the analysis for Xe+Au at 
50A MeV by: i) removing from $E_t$ all contributions from the specific 
$Z$ ($E_t^Z$) that we have selected (Fig. 5, middle panel). ii) by using 
only the $E_t$ of the light charge particles, $E_t^{LCP}$ (Fig. 5, right panel). 
In both cases, the Arrhenius plots remain linear for almost the entire range of 
$E_t$, and $\left<n_Z\right>$ changes by factors of 10 to 50. These results are 
similar to those obtained using the total $E_t$. We conclude that the 
linearity of the Arrhenius plots is not due to autocorrelation but to a 
thermal/statistical emission process dominated by phase space. 

\begin{figure}[tp]
\centerline{\psfig{file=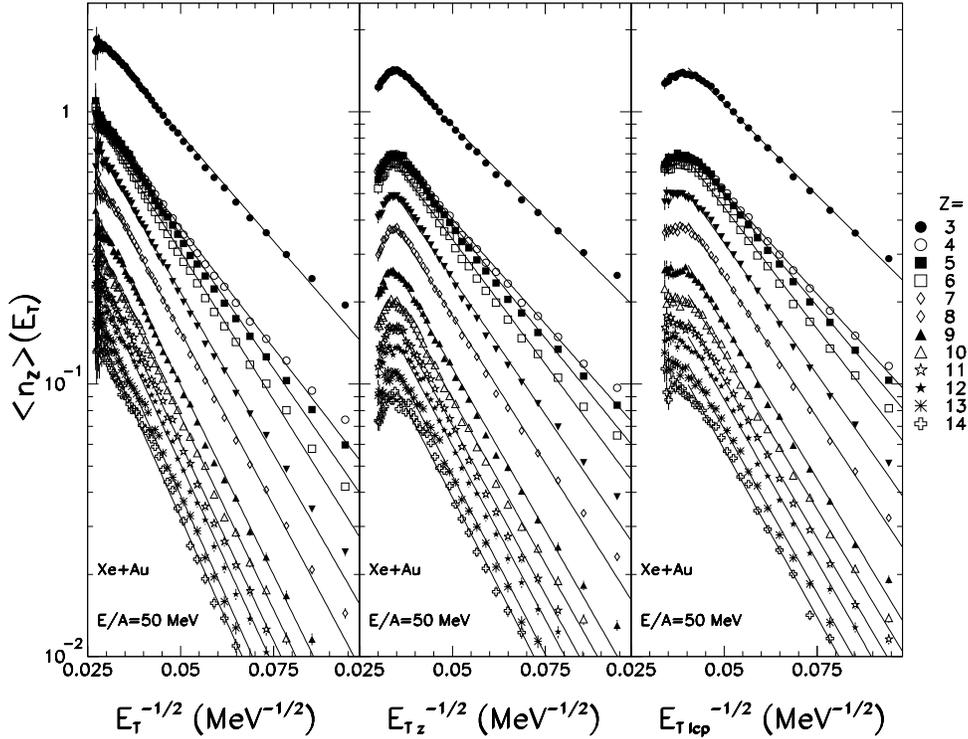,height=10.0cm,width=13.0cm}}
\caption{The average yield per event of different elements (symbols) 
as a function of $1/\protect\sqrt{E_t}$ for the reaction 
Xe+Au data at 50A MeV using the total transverse energy $E_t$ (left), the 
transverse energy of all charged particles excluding the Z that we have 
selected $E_t^Z$ (middle) and (right) that only of the light charged 
particles (LCPs) $E_t^{LCP}$. The lines are fits to the data using a 
Boltzmann form for $\left< n_Z \right>$.}
\end{figure}

We have observed experimentally that the maximum values of 
the new $E_t$ scales (either $E_t^Z$ or $E_t^{LCP}$) correspond to events 
in which fragments of a given $Z$ (or all IMFs) are absent.
Therefore, in our attempt to avoid autocorrelation by excluding from 
$E_t$ all IMFs ($E_t^{LCP}$) or the $Z$ value under investigation 
($E_t^Z$), we have introduced another kind of autocorrelation. For 
example, excluding from $E_t$ all fragments of charge $Z$ to produce 
$E_t^Z$ necessarily requires that for those events where $E_t^Z \approx E_t$, 
the yield $n_Z \rightarrow 0$. This produces the visible turnover of the 
Arrhenius plots in the bottom panels of Fig. 5 (the same argument also 
applies to $E_t^{LCP}$).

Finally, even though we have constructed the Arrhenius plots from three 
different $E_t$ scales, the slopes associated with these plots always become
steeper with increasing $Z$ values. This is what we would expect
if the slopes parameters are related to physical fragmentation
barriers. Moreover, the rate of change of the slopes with various
$E_t$ scale is the same. This is shown in Fig. 6 where the various
sets of barriers have been normalized to $Z$=6 from the full $E_t$
scale.

\begin{figure}[tp]
\centerline{\psfig{file=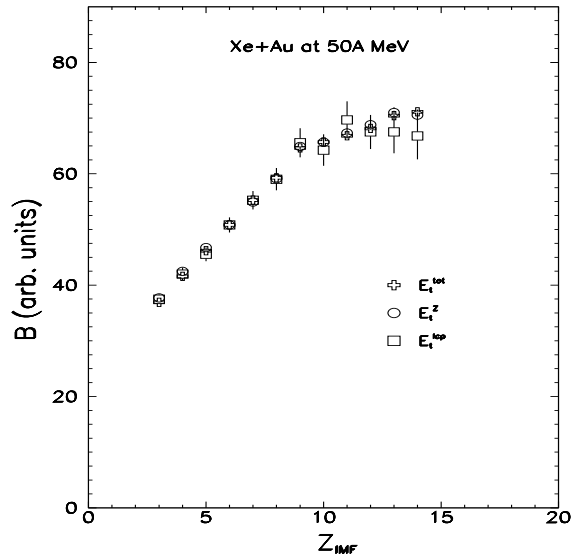,height=7.5cm,width=7.5cm}}
\caption{ Slopes of the Arrhenius plots, normalized to 
$Z$=6, for Xe+Au at 50A MeV as a function of $Z$ using the indicated 
definitions of $E_t$.}
\end{figure}

\subsection{Summary and Outlook}

In heavy ion reactions at intermediate energies, a complex dynamical 
behaviour is observed in the entrance channel. However, in order to 
understand the nature of the fragmentation process, one must rely 
on observables other than velocity plots, and their associated kinematic
variables.

The evolution of multiplicities of neutrons, light charged particles or IMF 
and of their corresponding transverse energies with $N_{IMF}$ or $E_t$ does 
not provide convincing evidence for the claim of a dynamical IMF production. 
In the first case~\refnote{\cite{Tok96}}, the behaviour is a rather general 
one and is found in any statistical model. In the second case~\refnote{\cite{Tok97}},
the anomalous features associated with dynamical IMF production are most 
likely due to an experimental artifact.

Armed with observables that have have been successful for low energy 
nuclear reactions, we have used the probabilities and excitation functions 
to probe the nature of the fragmentation process. The n-fold probabilities
of individual $Z$ values are shown to follow Poissonian
distributions, and as such, are reducible. The experimental observation 
of Poissonian reducibility means that IMF production is dominated by a 
stochastic process. Of course stochasticity falls directly in the realm 
of statistical decay. It is less clear how it would fare within the 
framework of a dynamical model without appealing for chaoticity
or ergodicity. Futhermore, the thermal scaling of $\left< n_Z \right>$ 
suggest that it has the Boltzmann form
\begin{equation}
\left< n_Z \right> \propto e^{-B_{Z}/T}
\end{equation}

It is important to recall that by considering individual $Z$ values, one
obtains Arrhenius plots free of distortion or autocorrelation.
Additionnally, this form permits the extraction of a fragmentation ``barrier'' 
$B_Z$ for each $Z$. The behaviour described in Eq. 2 is similar
to that observed in the fission process of ref.~\cite{Mor69}. The emission 
probability of a given $Z$ is controlled by its emission barrier and the 
temperature. 

\subsection{Acknowledgments}
This work was supported by the Director, Office of Energy Research, 
Office of High Energy and Nuclear Physics, 
Nuclear Physics Division of the US Department of Energy, 
under contract DE-AC03-76SF00098. One of us (L.B)
acknowledge a fellowship from the National Sciences and Engineering
Research Council (NSERC), Canada.


\begin{numbibliography}

\bibitem{Fis67}
M.E. Fisher{{,}}
\newblock {Physica {\bf 3}}, 225 (1967).

\bibitem{Stu92}
D. Stuffer and A. Aharony{{,}}
\newblock Introduction to percolation theory, 2nd Ed. (Taylor and Francis,
London, 1992) pp.181.

\bibitem{Fin82}
J.E. Finn {\em et al.{,}}
\newblock {Phys. Rev. Lett {\bf 49}}, 1321 (1982).

\bibitem{Sie83}
J.P. Siemens{{,}}
\newblock {Nature {\bf 305}}, 410 (1983).

\bibitem{Pan84}
A.D. Panagiotou {\em et al.{,}}
\newblock {Phys. Rev. Lett {\bf 52}}, 496 (1984).

\bibitem{Bor92} B. Borderie, Ann, de Phys. {\bf 17}, 349 (1992).

\bibitem{Mor93} L.G. Moretto and G.J. Wozniak, Ann. Rev. Nucl. Part. Sci. 
{\bf 43}, 379 (1993).


\bibitem{Des93}
P.~D\'esesquelles {\em et al.{,}}
\newblock {Phys. Rev. C {\bf 48}}, 1828 (1993).

\bibitem{Kre93}
P.~Kreutz {\em et al.{,}}
\newblock {Nucl. Phys. A{\bf 556}}, 672 (1993).

\bibitem{Gil94}
M.L. Gilkes {\em et al.{,}}
\newblock {Phys. Rev. Lett {\bf 73}}, 1590 (1994).

\bibitem{Poc95}
J. Pochodzalla {\em et al.{,}}
\newblock {Phys. Rev. Lett {\bf 75}}, 1040 (1995).

\bibitem{Ben95}
J. Benlliure{,}
\newblock Ph.D. thesis, University of Valencia, Spain, 1995 (unpublished).

\bibitem{Bea96c}
L. Beaulieu{{,}}
\newblock Ph.D. thesis, Universit'e Laval, Canada, 1996 (unpublished).

\bibitem{Mas96}
P.F. Mastinu {{\em et al,}}
\newblock Phys. Rev. Lett. {\bf{ 76}}, 2646 (1996).


\bibitem{Bea96b}
L. Beaulieu {\em et al.{,}}
\newblock {Phys. Rev. C {\bf 54}}, R973 (1996).

\bibitem{Sch96}
A. Sch\"uttauf {\em et al.{,}}
\newblock {Nucl. Phys. A {\bf 607}}, 457 (1996).

\bibitem{Poc97}
J. Pochodzalla{,}
\newblock {Prog. Part. Nucl. Phys. {\bf 39}}, 443 (1997).

\bibitem{Hau98}
J.A. Hauger {\em et al.{,}}
\newblock {Phys. Rev. C {\bf 57}}, 764 (1998).


\bibitem{Lot92}
B. Lott {\em et al.{,}}
\newblock {Phys. Rev. Lett. {\bf 68}}, 3141 (1992).

\bibitem{Que93}
B.M. Quednau {\em et al.{,}}
\newblock {Phys. Lett. {\bf B309}}, 10 (1993).

\bibitem{Lec94}
J.F. Lecolley {\em et al.{,}}
\newblock {Phys. Lett. {\bf B325}}, 317 (1994).

\bibitem{Pet95}
J. P\'eter {\em et al.{,}}
\newblock {Nucl. Phys. {\bf A593}}, 95 (1995).

\bibitem{Bea96}
L. Beaulieu {\em et al.{,}}
\newblock {Phys. Rev. Lett. {\bf 77}}, 462 (1996).

\bibitem{Pha95} L. Phair et al., Phys. Rev. Lett. {\bf 75}, 213 (1995).

\bibitem{Mor95b} L.G. Moretto et al., Phys. Rev. Lett. {\bf 76}, 372
(1996).


\bibitem{Monto94} C.P. Montoya {\em et al.}, Phys. Rev. Lett {\bf 73}, 
3070 (1994).

\bibitem{Luka97} J. Lukasik {\em et al.}, Phys. Rev. C {\bf 55}, 
1906 (1997).

\bibitem{Laro97} Y. Larochelle {\em et al.}, Phys. Rev. C {\bf 55}, 
1869 (1997).

\bibitem{Tok95} J. Toke {\em et al.}, Phys. Rev. Lett. {\bf 75},
2920 (1995).

\bibitem{Leco95} J.F. Lecolley {\em et al.}, Phys. Lett. B {\bf 354},
202 (1995).

\bibitem{Demp96} J.F. Dempsey {\em et al.}, Phys. Rev. C {\bf 54},
1710 (1996).

\bibitem{Tok96} J. Toke {\em et al.}, Phys. Rev. Lett {\bf 77},
3514 (1996).

\bibitem{Tok97} J. Toke {\em et al.}, Phys. Rev. C {\bf 56},
R1683 (1997).

\bibitem{Mor95} L.G. Moretto {\em et al.}, Phys. Rev. Lett. 
{\bf 74}, 1530 (1995).

\bibitem{Tso95} K. Tso et al., Phys. Lett. B {\bf 361}, 25 (1995).

\bibitem{Mor97} L.G. Moretto, {\em et al.}, Phys. Rep. {\bf 287}, 249
(1997).

\bibitem{Pha96} L. Phair {\em et al.}, Phys. Rev. Lett {\bf 77}, 
822 (1996). 

\bibitem{Bea98} L. Beaulieu {\em et al.}, Submitted to Phys. Rev. Lett. 

\bibitem{Mor93b} L.G. Moretto, {\em et al.}, Phys. Rev. Lett. {\bf 71}, 3935
(1993).



\bibitem{Tsa97} M.B. Tsang {\em et al.}, Phys. Rev. Lett. {\bf 80}, 
1178 (1998)

\bibitem{Sku97} W. Skulski {\em et al.}, to appear in 
Proc. 13th Workshop on Nuclear Dynamics, Key West, Florida (1997).

\bibitem{Sou92} R.T. de Souza {\em et al.}, Nucl. Inst. Meth. {\bf A 311},
109 (1992).

\bibitem{Keh92} W.C. Kehoe {\em et al.}, Nucl. Inst. Meth. {\bf A 311},
258 (1992).


\bibitem{Bon95} J.P. Bondorf {\em et al.}, Phys. Rep. {\bf 257},
133 (1995).

\bibitem{Pha98a} L. Phair {\em et al.}, Accepted in Phys. Rev. Lett..

\bibitem{Col98} N. Colonna, private communication.

\bibitem{Pha98b} L. Phair {\em et al.}, to be published.

\bibitem{Str90} D.W. Stracener {\em et al.}, Nucl. Inst. Meth. {\bf A 294},
485 (1990).

\bibitem{Mor69} L.G. Moretto, Phys. Rev. {\bf 179}, 1176 (1969).

\end{numbibliography}
\end{document}